\documentclass[preprint,12pt]{elsarticle}

\usepackage[a4paper, total={6in, 8in}]{geometry}

\usepackage{amssymb}
\usepackage{float}
\usepackage{xcolor}

\journal{}

\begin{document}

\begin{frontmatter}



\title{Adhesive contact mechanics of bio-inspired pillars: exploring hysteresis and detachment modes}


\author[inst1]{G. Violano}

\affiliation[inst1]{organization={Department of Mechanics, Mathematics and Management,Polytechnic University of Bari},
            addressline={Via E. Orabona 4}, 
            city={Bari},
            postcode={70125},
            country={Italy}}

\author[inst1]{S. Dibitonto}
\author[inst1]{L. Afferrante}

\begin{abstract}
Engineering technologies frequently draw inspiration from nature, as exemplified in bio-inspired adhesive surfaces. These surfaces present textures adorned by pillars, mimicking the topography found on the pads of certain animals renowned for their exceptional adhesive capabilities. The adhesive response is strongly influenced by the morphology of these pillars. \textcolor{black}{In typical existing models, perfect bonding conditions are assumed between the pillar and the countersurface, and solely the detachment process of the pillar from the countersurface is investigated.}

\textcolor{black}{The proposed model, based on the assumption that interactions at the interface are governed by van der Waals forces modeled by the Lennard-Jones potential law, enables the examination of the entire approach and retraction cycle, tracking the movement of the pillar towards and away from the countersurface.}

\textcolor{black}{Our findings reveal that adhesive contact mechanics is primarily influenced by the geometry of the pillar and the potential presence of interfacial 'defects', which in turn affect the distribution of contact pressure. Furthermore, we show that the detachment process may simultaneously involve various modes of separation, such as crack propagation from outer edge, crack propagation from inner defects, and uniform decohesion. This suggests that existing theoretical models alone cannot fully elucidate the complexity of detachment phenomena. Additionally, we anticipate the occurrence of hysteretic losses during the approach-retraction cycle, attributed to pull-in and pull-off contact jumps. Adhesive hysteresis is a phenomenon consistently observed in experiments but frequently overlooked in existing models.}
\end{abstract}


\end{frontmatter}


\section{Introduction}
Bio-inspired adhesives are engineered to replicate the skills found in specific animal species \cite{brodoceanu2016,favi2014,purtov2015}, whose remarkable adhesive performance can be attributed to two primary mechanisms. In the gecko's pad, for example, adhesion is guaranteed by its fibrillar hierarchical structure \cite{autumn2002}. Other species, like some beetles, achieve climbing capacity through the distinctive microstructures on their pads \cite{gorb2007}. In both cases, we are dealing with "dry" adhesion, which is solely attributed to the van der Waals forces operating at the contact interface \cite{autumn2002B,hosoda2021}. Artificial adhesives are frequently produced with soft polymers and present a surface texture covered with micropillars, similar to those found on beetle pads \cite{zhang2015,lee2018,hensel2018}. The typical shapes of these pillars include the cylindrical one and the mushroom-shaped geometry. Both experimental evidence and theoretical studies \cite{gorb2007,bullock2011} have established that the incorporation of a cap extending the punch diameter in mushroom-shaped pillars significantly enhances adhesive performance.

\begin{figure}[H]
\centering\includegraphics[width=\textwidth]{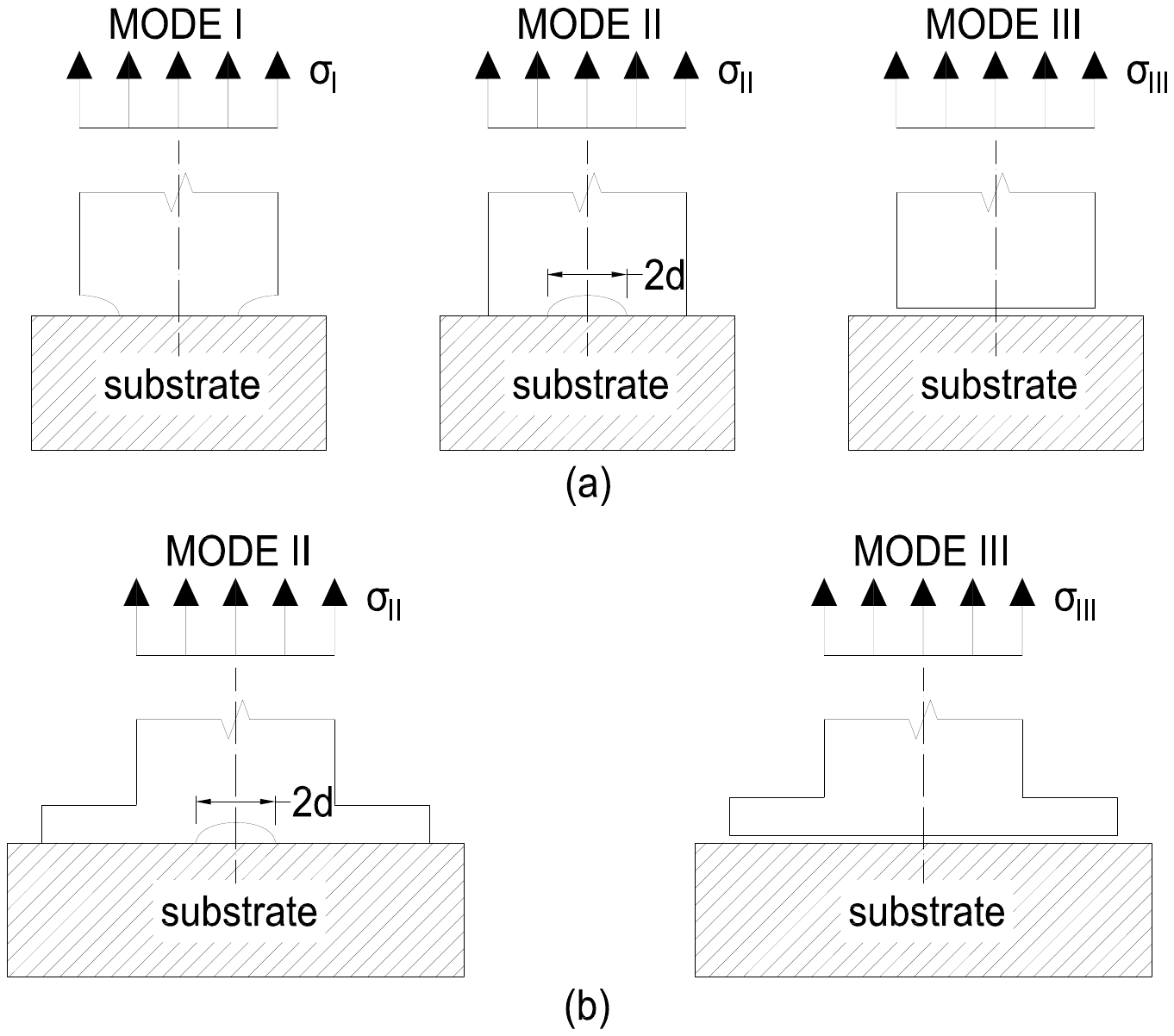}
    \caption{Detachment modes for a) cylindrical pillars and b) mushroom-shaped pillars. }
    \label{Modes}
\end{figure}

In the last years, several studies have been conducted on the influence of pillars geometry \cite{kim2020, carbone2013, aksak2014, afferrante2013}, aspect ratio \cite{micciche2014,paretkar2013,greiner2007}, material properties \cite{gorb2007,wang2014,fischer2017}, and surfaces roughness of the countersurface \cite{kasem2013,afferrante2023A,afferrante2023B} on the adhesive performance of microstructured adhesives.

Carbone et al. \cite{carbone2011} performed a theoretical study on the detachment of a soft elastic pillar from a rigid flat surface, highlighting that the debonding mechanisms are affected by the specific contact geometry. Also, they classified the potential debonding modes as sketched in Fig. \ref{Modes}. In the case of cylindrical pillars, adhesive bond rupture occurs due to crack propagation starting from the external perimeter and advancing toward the inner region. The presence of the cap in mushroom-shaped pillars, inhibits crack formation at the plate's edge. 

Zhang et al. \cite{zhang2021} performed experiments and Finite Element (FE) analyses to study the detachment of an isolated mushroom-shaped pillar. They observed that the geometry of the mushroom cap strongly affects the interfacial crack behavior and the pull-off stress, finding the existence of an optimum cap thickness that maximizes the adhesion force in agreement with results reported in Carbone and Pierro \cite{CarboneSMALL}.

The above cited works demonstrate that the detachment mode depends on the pressure distribution at the contact interface. However, the examination of the pressure distribution is challenging when relying solely on analytical models and presents obvious difficulties in terms of direct observation through experiments. Numerical models are a valuable tool for tracking the pressures at the interface, providing insights into their relationship with the detachment mode \cite{zhang2021}.

\textcolor{black}{Several numerical approaches have been proposed to study adhesive phenomena between an elastic body and a rigid indenter \cite{rey2017,salehani2019,popov2021,sanner2022}. However, many of these studies, such as those based on the Boundary Element Method (BEM), are limited by the assumption that the deformable body is of infinite size, like an elastic half-space or an infinitely long elastic layer. This assumption holds true only when the contact area is small compared to the overall size of the bodies in contact.
In contrast, the versatility of the  Finite Element Method (FEM) allows assigning deformable properties to finite-sized bodies with complex geometries, such as mushroom-shaped pillars. Therefore, the model originally developed in Refs. \cite{afferrante2022, violano2022A} is refined here to investigate the mechanism of detachment of soft cylindrical and mushroom-shaped pillars from a rigid flat countersurface. This refinement allows for a more precise analysis of the interaction mechanics, considering the finite size and geometry of the pillars, which are crucial for accurately predicting adhesion behavior in practical applications.}

\textcolor{black}{The present approach models van der Waals interactions at the interface using the Lennard-Jones potential law. The implementation of a finite potential is a much more realistic condition compared to the assumption of infinitely short-range adhesion, which is typical of models based on energetic approaches \cite{carbone2013,carbone2011}. Furthermore, while theoretical models can predict detachment modes based on energetic equilibrium \cite{afferrante2013}, our model provides the complete pressure distribution at the contact interface. This detailed information accurately reveals the exact detachment mechanisms that occur, offering a more comprehensive understanding of the interaction dynamics.}

\textcolor{black}{In contrast to prior theoretical and numerical models \cite{carbone2013,aksak2014,zhang2021}, our approach does not require the assumption of initial bonded conditions between the pillar and the countersurface. Instead, it allows for comprehensive modeling of both the approach and retraction processes, facilitating investigation of the non-conservative behavior associated with adhesive hysteresis, a phenomenon typically observed in experimental studies \cite{zhang2021}.}

Results show that the detachment mode is governed by the interfacial pressure distribution, which can be adjusted by i) changing the contact geometry of the pillar and/or ii) inserting an interfacial defect \cite{carbone2011}. In real applications, the presence of a defect at the interface can be accidental, as, for example, due to the deposition of dust particles, interfacial air entrapment, or geometrical imperfections of the pillar \cite{carbone2012}. On the other hand, the presence of a "defect" can be designed by changing the mechanical and adhesive properties of portions of the pillar \cite{kossa2023}. \textcolor{black}{Our findings reveal that the detachment process may involve multiple modes of separation, and hysteretic losses originate from pull-in and pull-off contact jumps occurring during the loading and unloading phases, respectively.}

\section{Theoretical background}
In this section, we provide a brief overview of the detachment modes of an elastic pillar from a rigid countersurface, as detailed in Ref. \cite{carbone2011}.

Figure \ref{Modes}a shows the three possible detachment modes for a cylindrical pillar: Mode I corresponds to crack propagation from the contact edge, Mode II corresponds to the propagation of an interfacial defect, and Mode III corresponds to the decohesion due to the achievement of theoretical contact strength. 

Mode I occurs when the average contact pressure at the interface exceeds the value

\begin{equation}
    \sigma_{\mathrm{I}}= \sqrt{\frac{8E^{*}\Delta\gamma}{\pi R}}
    \label{sigma1}
\end{equation}
being $\Delta \gamma $ the surface energy of adhesion, $E^{*}= E/(1-\nu^2)$ the reduced elastic modulus of the pillar and $R$ its radius. With $E$ and $\nu$ we denoted the Young's modulus and the Poisson ratio, respectively. \textcolor{black}{Equation (\ref{sigma1}) represents a well-established result in contact mechanics, as demonstrated in, for instance, Ref. \cite{maugis2000}.}

Mode II occurs when a central defect with length $2d$ is located at the contact interface and the average contact pressure exceeds the value

\begin{equation}
    \sigma_{\mathrm{II}}= \sqrt{\frac{\pi E^{*}\Delta\gamma}{2 d}}.
     \label{sigma2}
\end{equation}
\textcolor{black}{The estimate of $\sigma_{\mathrm{II}}$ has been derived in Ref. \cite{carbone2011}. Here, we recall that the change in the total energy necessary to generate a defect of radius $d$ is estimated as $\Delta U \approx -4 \sigma^{2} d^{3}/(3 E^{*})+\pi d^{2} \Delta \gamma$, being $\sigma$ the average stress at the interface. Given the applied stress, the change in total energy reaches a maximum value, namely the energy barrier against propagation, when the defect is $d_{\mathrm{II}} = \pi \Delta \gamma E^{*}/(2 \sigma^{2})$. Forcing the conditions $d=d_{\mathrm{II}}$ and $\partial{\Delta U}/\partial{d}=0$, one can derive the critical stress $\sigma_{\mathrm{II}}$.}

Finally, Mode III occurs at the theoretical contact stress
\begin{equation}
    \sigma_{\mathrm{III}}= \frac{\Delta\gamma}{\epsilon}
     \label{sigma3}
\end{equation}
being $\epsilon$ the range of action of van der Waals forces. Only the mode corresponding to the minimum value among the critical stresses $\sigma_{\mathrm{I}}$, $\sigma_{\mathrm{II}}$ and $\sigma_{\mathrm{III}}$ takes place.

 When a plate is added at the top of the pillar, it is typically designed to eliminate the stress singularity at the outer edge, thereby preventing the occurrence of Mode I detachment \cite{carbone2011}. As a result, the remaining detachment mechanisms are Mode II and Mode III (see figure \ref{Modes}b).

The above theoretical results have been derived under the ideal assumption of infinitely short-range adhesion, characterized by strong interactions acting within the contact area \cite{johnson1971,violano2019}, a common trait in soft matter systems \cite{lorenz2013}. However, such theoretical model do not provide information about the hysteretic loading-unloading behavior of soft pillars, notwithstanding the adhesive hysteresis is often observed in soft matter contacts \cite{violano2021rateA,violano2019B}.

\section{Formulation of the problem}
\begin{figure}[H]
    \centering\includegraphics[width=\textwidth]{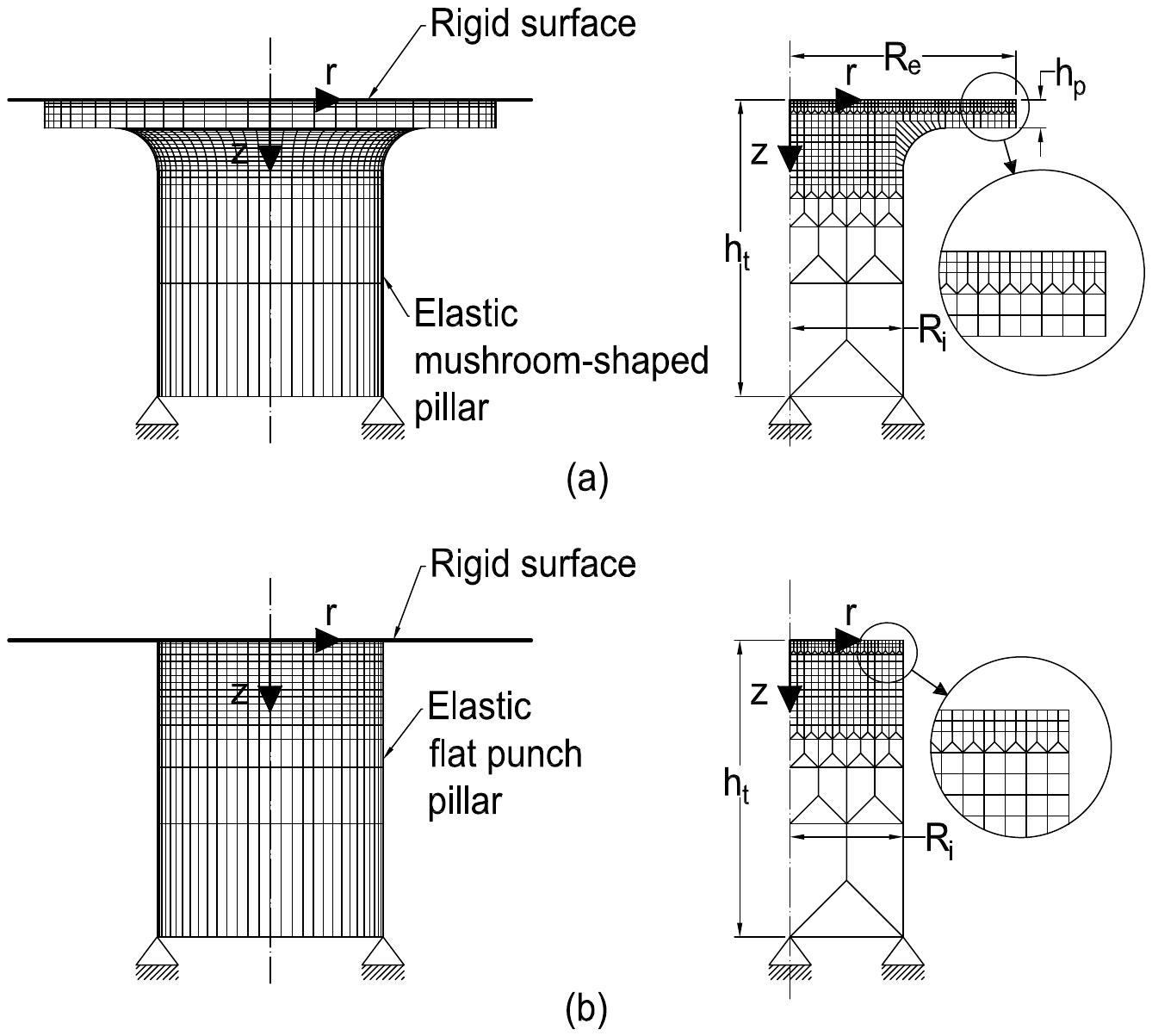}
    \caption{The geometry of the problem under investigation: adhesive contact between a rigid flat surface and an elastic mushroom-shaped pillar (a) and a cylindrical pillar (b). The geometrical parameters showed in the figure have the following meaning: $R_{\mathrm{i}}$ and $h_{\mathrm{t}}$ are the radius and height of the pillar, $R_{\mathrm{e}}$ and $h_{\mathrm{p}}$ are the radius and height of the plate.}
    \label{Fem}
\end{figure}

\textcolor{black}{The problem under investigation is depicted in Fig. \ref{Fem}, where an elastic axisymmetric pillar is pressed into contact against a rigid countersurface and subsequently pulled apart from it.}

\textcolor{black}{Unless otherwise specified, we have assumed $\Delta\gamma/(E^{*} \epsilon) = 0.02$, with $\epsilon =1$ nm. Such values are typical of van der Waals adhesive interactions \cite{PR2014}. All the parameters describing the pillar geometry are normalized with respect to the inner radius $R_{\mathrm{i}}$. The cylindrical pillar has total height $\hat{h}_\mathrm{t} = h_\mathrm{t}/R_{\mathrm{i}} = 2.38$, with $R_{\mathrm{i}} = 0.5$ $\mathrm{\mu}$m. The adopted geometry and aspect ratio are consistent with values taken from real textured adhesive surfaces \cite{aksak2011RATIO,delcampo2007ALTEZZA}. Additionally, for the mushroom-shaped pillar, we considered an external radius $\hat{R}_{\mathrm{e}}=2$ and a plate thickness $\hat{h}_\mathrm{p} = 0.22$; the radius of connection between the pillar and the plate is $\hat{\rho}\approx0.22$.}

\textcolor{black}{\subsection{Finite Element Model}}

\textcolor{black}{The contact problem is solved by utilizing the finite element approach described in Refs. \cite{afferrante2022, violano2022A}, where additional details can be found.}

\textcolor{black}{The rigid flat surface is modeled by its nodes, and employing a single master node to enforce force and displacement. The degrees of freedom of this master node are interconnected with those of all other nodes through constraint equations.}

\textcolor{black}{Due to the configuration's symmetry, we can analyze the problem in two dimensions using axisymmetric isoparametric elements with linear shape functions to model the pillar, which is constrained at its base. A regular mesh is employed for the pillar, with denser mesh at the contact interface, where the discretization is $\Delta x = R_{\mathrm{i}}/512$.}

To simulate the adhesive interactions between the pillar and the countersurface, nonlinear spring elements are placed at the contact interface. Their stiffness is determined based on a traction-displacement relation derived from the Lennard-Jones (LJ) potential law

\begin{equation}
\sigma_{\mathrm{LJ}} \left( r\right) =\frac{8\Delta \gamma }{3\epsilon }\left[ \left( 
\frac{\epsilon }{g\left( r\right) }\right) ^{3}-\left( \frac{\epsilon 
}{g\left( r\right) }\right) ^{9}\right]
\end{equation}%
where $g\left( r\right) $ is the interfacial gap, \textcolor{black}{being $r$ the radial coordinate (see figure \ref{Fem}). We stress that for the surface potential assumed in this study, the theoretical contact stress is $\sigma_{\mathrm{III}}\approx 1.023 \Delta \gamma /\epsilon$, but for simplicity we will adopt the value given in eq. (\ref{sigma3}).}

\textcolor{black}{The contact cycles during approach and retraction are simulated by controlling the displacement of the master node on the rigid flat surface. Consequently, the total contact force is calculated as the reaction force to the applied displacement. This value matches the sum of the forces due to the deformation of each individual contact element. This method was initially proposed by Muller et al. \cite{muller1983}, who used numerical summation of interactions to determine the force between a smooth sphere and a flat surface.}

\textcolor{black}{The presence of an internal defect with length $2d$ is simulated by removing the contact springs located from the center of the pillar to the horizontal coordinate $r=d$. From a realistic perspective, such a condition could be reached with pillars having mechanical and adhesive properties that vary radially \cite{kossa2023}.}

\textcolor{black}{During the approach and retraction of the indenter, jump-in and jump-off contact instabilities may occur due to adhesive forces, potentially causing numerical convergence issues. To ensure numerical stability, dashpot elements are added at the interface which activate only when unstable jumps occur. This approach stabilizes the simulation by smoothing out numerical oscillations and discontinuities.}

\section{Results}

In the presentation of the results, we shall refer to the following adimensional quantities: load $\hat{F}=F/(E^{*}R_{\mathrm{i}}^{2})$, contact radius $\hat{a}=a/R_{\mathrm{i}}$, penetration $\hat{\delta}=\delta/\epsilon$, pressure $\hat{\sigma} = \sigma \epsilon/\Delta\gamma$, size of the interfacial defect $\hat{d}=d/R_{\mathrm{i}}$.

\begin{figure}[H]
    \centering\includegraphics[width=\textwidth]{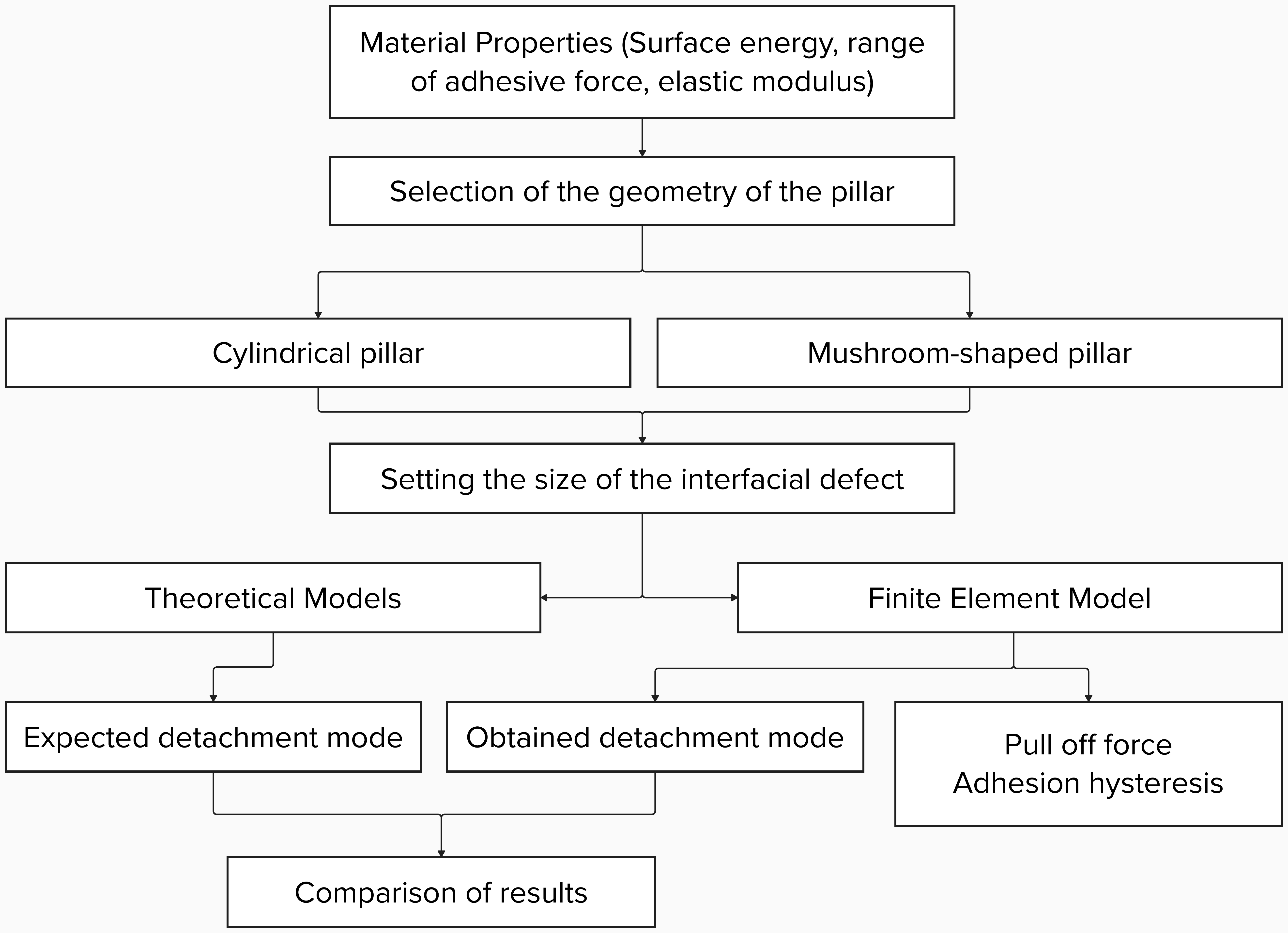}
    \caption{Flowchart of the present study.}
    \label{flow}
\end{figure}

\textcolor{black}{The flowchart presented in Figure \ref{flow} outlines the methodology adopted in this study. It begins with the selection of material properties, followed by the decision regarding the pillar geometry, either cylindrical or mushroom-shaped. Subsequently, the size of the inner interfacial defect for the pillar is determined. The next step entails the utilization of theoretical models outlined in Section 2 to predict the detachment mode; this prediction is compared with the outcomes derived from our numerical simulations. Furthermore, the numerical model is used for examining the loading-unloading process of the adhesive pillar, computing both the pull-off force and the hysteretic energy loss.}

\subsection{Cylindrical Pillar - Mode I}
In absence of defect at the contact interface ($\hat{d}=0$), from eqs. (\ref{sigma1}), (\ref{sigma2}), and (\ref{sigma3}), we can derive the following values for the dimensionless critical stresses
\begin{equation}
   	 \hat{\sigma}_{\mathrm{I}}= 0.505 \quad \hat{\sigma}_{\mathrm{II}}\to \infty \quad \hat{\sigma}_{\mathrm{III}}=1,
    \label{sigmaFPmode1}
\end{equation}
and, hence, anticipate the occurrence of Mode I detachment.

\begin{figure}[H]
    \centering\includegraphics[width=\textwidth]{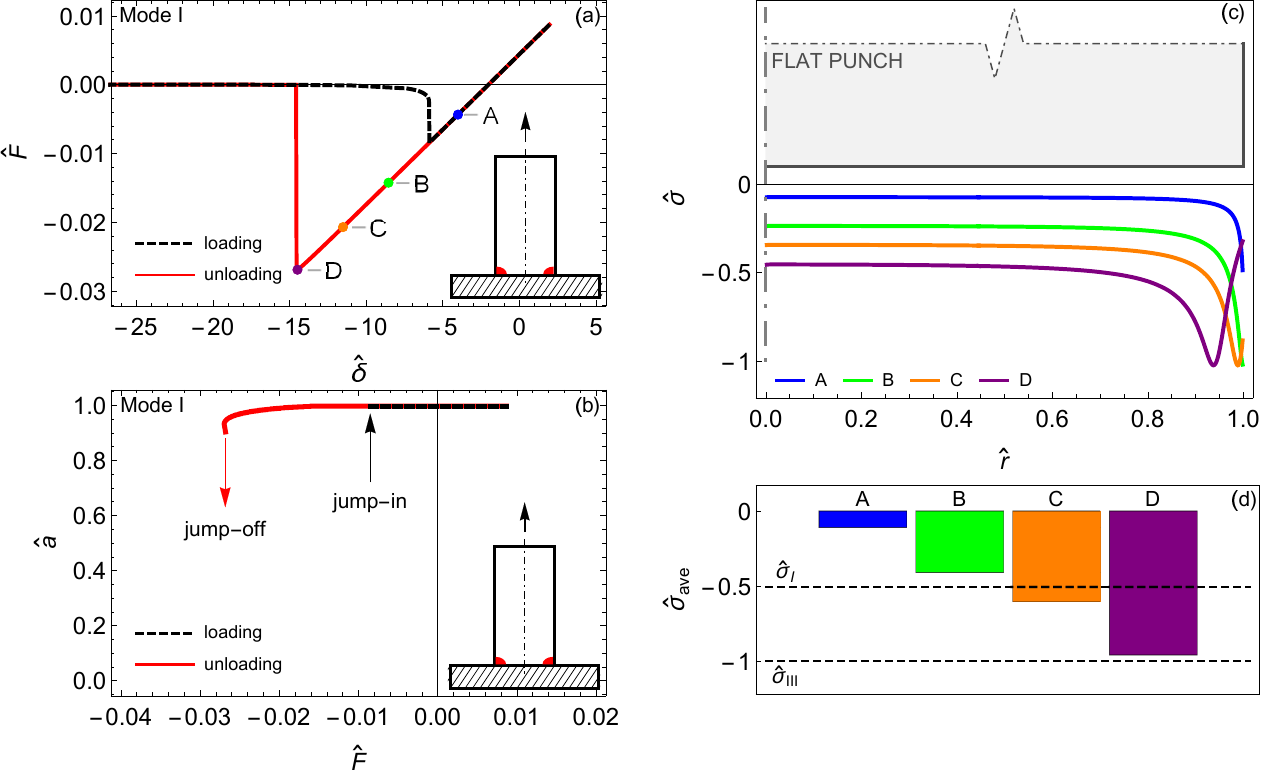}
    \caption{Mode I detachment of the cylindrical pillar: a) Normalized force $\hat{F}$ plotted against the dimensionless approach $\hat{\delta}$ during the loading (black dashed line) and unloading (red solid line) phases. b) Dimensionless contact radius $\hat{a}$ shown as a function of the force $\hat{F}$. c) Interfacial pressure distribution $\hat{\sigma}$ and d) average pressure $\hat{\sigma}_\mathrm{ave}$ displayed for various values of the imposed approach $\hat{\delta}$.}
    \label{FPmode1}
\end{figure}

Figure \ref{FPmode1}a illustrates the dimensionless force $\hat{F}$ as a function of the imposed dimensionless approach $\hat{\delta}$, during both the loading (black dashed line) and retraction (red solid line) phases. The pull-off force (point D) attained during retraction serves as a direct measure of adhesion strength. The region between the approach and retraction paths indicates the magnitude of hysteretic loss within a loading-unloading cycle. \textcolor{black}{This phenomenon is due to adhesive hysteresis; as the pillar approaches the flat surface, adhesive forces cause the contact to form abruptly at a critical jump-in distance. When the pillar is retracted, these same adhesive forces resist the separation, causing the contact to break at a larger jump-off distance. This difference in distances creates a hysteresis loop, representing energy loss in the system. Physically, this behavior is due to the nature of adhesive forces at the interface, which create a stronger attraction when forming contact and a stronger resistance when breaking contact. This results in a non-conservative process, where some of the energy applied during loading is dissipated, typically as heat, during unloading \cite{maugis2000,violano2021HYST,wang2021}.}

Pull-in and pull-off jumps are also depicted in Fig. 3b, where the contact radius $\hat{a}$ is given in terms of the applied load. Defining the contact radius rigorously can be a matter of debate, given that the Lennard-Jones force law requires a non-zero gap between contacting surfaces \cite{afferrante2022,feng2001}.
In our calculations, we define the contact radius as the sum of segments where the interfacial gap is below a specified threshold. Specifically, contact is assumed to occur when $(g(r)-\epsilon)/\epsilon < 0.1$. We validated this criterion in the case of classical JKR contact \cite{johnson1971} between a rigid sphere and an elastic half-space.

Figure \ref{FPmode1}c displays the interfacial pressure distribution corresponding to four different values of the approach, denoted by points A, B, C, and D in Fig. \ref{FPmode1}a. At point A, the contact pressure exhibits its maximum value at the outer edge of the contact ($\hat{r} = r/R_{\mathrm{i}}=1$). As retraction progresses, the pressure at the edge grows until it reaches a maximum value (point B), equal to $\hat{\sigma}_{\mathrm{III}}$; for further reductions of the penetration $\hat{\delta}$, the average pressure $\hat{\sigma}_{\mathrm{ave}}=F\epsilon/(\pi a^{2}\Delta\gamma)$ approaches the critical stress $\hat{\sigma}_{\mathrm{I}}$ (see Fig. \ref{FPmode1}d), and the maximum pressure moves from the edge towards the center of the pillar (points C and D).

This behavior is characteristic of Mode I detachment, wherein an interfacial crack initially forms at the outer edge of the contact and then propagates toward the internal region. When reaching pull-off, the average pressure closely approaches the critical stress $\hat{\sigma}_{\mathrm{III}}$ (see Fig. \ref{FPmode1}d), leading to the loss of contact.

\subsection{Cylindrical Pillar - Mode I + Mode II}
The presence of an interfacial "defect" of radius $\hat{d}=0.446$, leads to the following values of the critical stresses 
\begin{equation}
   	 \hat{\sigma}_{\mathrm{I}}= 0.505 \quad \hat{\sigma}_{\mathrm{II}}= 0.593 \quad \hat{\sigma}_{\mathrm{III}}=1.
    \label{sigmaFPmode1mode2}
\end{equation}
Theoretical calculations \cite{carbone2011} predict detachment should occurs according to Mode I.

\begin{figure}[H]
    \centering\includegraphics[width=\textwidth]{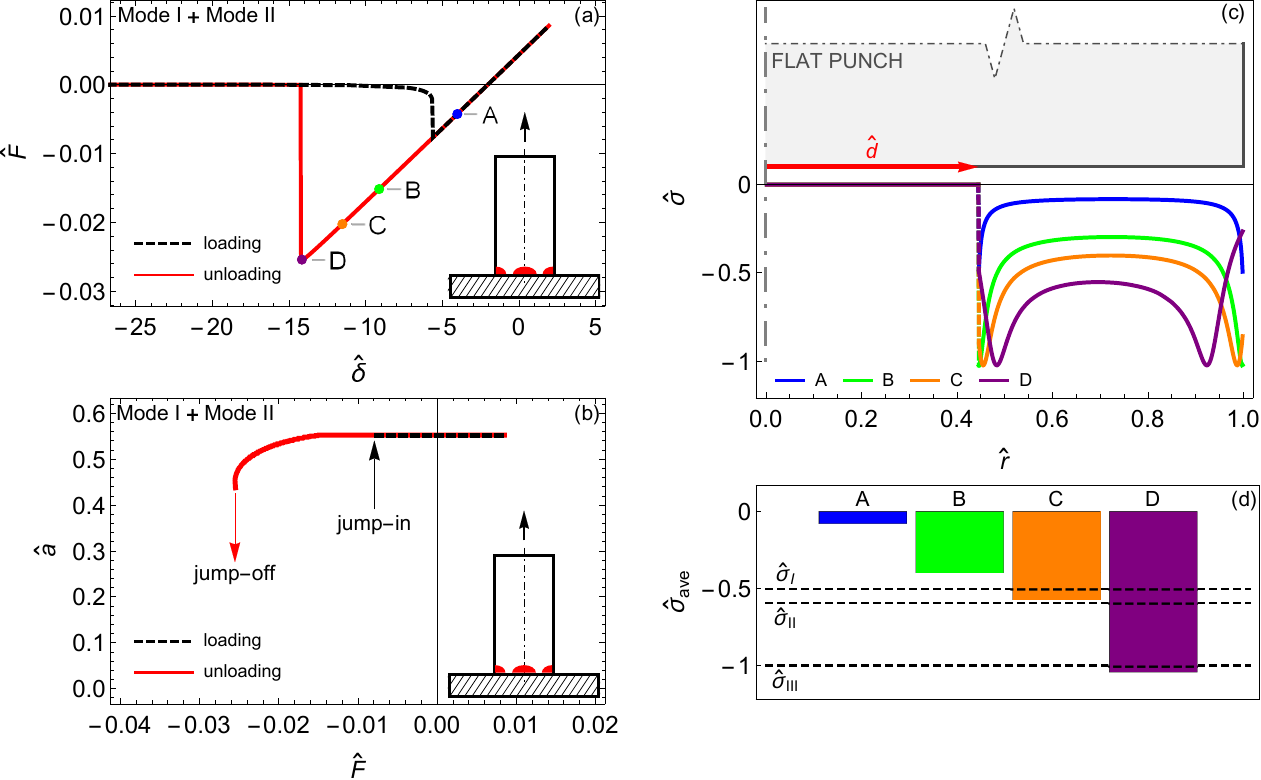}
    \caption{Mode I + Mode II detachment of the cylindrical pillar: a) Normalized force $\hat{F}$ plotted against the dimensionless approach $\hat{\delta}$ during the loading (black dashed line) and unloading (red solid line) phases. b) Dimensionless contact radius $\hat{a}$ shown as a function of the force $\hat{F}$. c) Interfacial pressure distribution $\hat{\sigma}$ and d) average pressure $\hat{\sigma}_\mathrm{ave}$ displayed for various values of the imposed approach $\hat{\delta}$.}
    \label{FPmode1mode2}
\end{figure}

Figures \ref{FPmode1mode2}a and \ref{FPmode1mode2}b show the curves $\hat{F}(\hat{\delta})$ and $\hat{a}(\hat{F})$, respectively. No significant difference can be found compared to the counterparts of Fig. \ref{FPmode1}.
On the contrary, a significant difference in the pressure distribution at the contact interface (Fig. \ref{FPmode1mode2}c) is evident. Before the detachment is triggered (curve A), the pressure exhibits two peaks - one at the external edge of the pillar and the other at the edge of the defect. As the approach $\hat{\delta}$ decreases, two detachment modes become apparent: when $\hat{\sigma}_{\mathrm{ave}}$ approaches $\hat{\sigma}_{\mathrm{I}}$, the pressure peak at the outer edge of the pillar reaches its maximum value (curve B), activating Mode I detachment; a further decrease in $\hat{\delta}$ leads to $\hat{\sigma}_{\mathrm{ave}}$ approaching $\hat{\sigma}_{\mathrm{II}}$, causing the pressure peak at the defect edge to reach its maximum value (curve C), triggering contact debonding even from the defect edge (Mode II detachment).
Finally, at point D (pull-off), when the average pressure $\hat{\sigma}_{\mathrm{ave}}$ equals the critical stress $\hat{\sigma}_{\mathrm{III}}$, the complete contact rupture occurs.

If the defect radius is increased ($\hat{d}=0.78$), the resulting critical stresses become:
\begin{equation}
   	 \hat{\sigma}_{\mathrm{I}}= 0.505 \quad \hat{\sigma}_{\mathrm{II}}= 0.449 \quad \hat{\sigma}_{\mathrm{III}}=1,
    \label{sigmaFPmode1mode2}
\end{equation}
and detachment is expected to follow Mode II.

\begin{figure}[H]
    \centering\includegraphics[width=\textwidth]{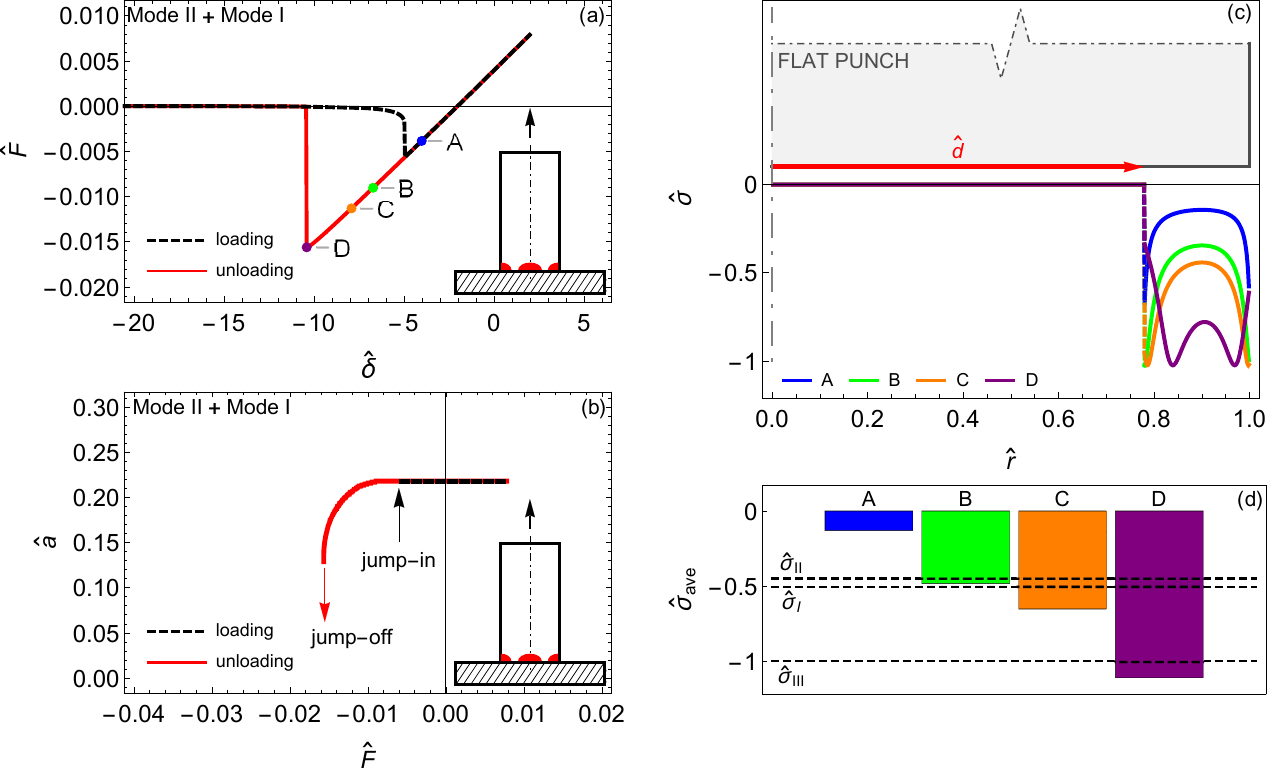}
    \caption{Mode II + Mode I detachment of the cylindrical pillar: a) Normalized force $\hat{F}$ plotted against the dimensionless approach $\hat{\delta}$ during the loading (black dashed line) and unloading (red solid line) phases. b) Dimensionless contact radius $\hat{a}$ shown as a function of the force $\hat{F}$. c) Interfacial pressure distribution $\hat{\sigma}$ and d) average pressure $\hat{\sigma}_\mathrm{ave}$ displayed for various values of the imposed approach $\hat{\delta}$.}
    \label{FPmode2mode1}
\end{figure}

The curves $\hat{F}(\hat{\delta})$ and $\hat{a}(\hat{F})$ are depicted in Figs. \ref{FPmode2mode1}a and \ref{FPmode2mode1}b, respectively. 
In this case, we can observe a behavior similar to the one described earlier, with the distinction that the contact debonding initially occurs from the defect edge when $\hat{\sigma}_{\mathrm{ave}}$ approaches $\hat{\sigma}_{\mathrm{II}}$ (curve B in Fig. \ref{FPmode2mode1}c).
When $\hat{\sigma}_{\mathrm{ave}}$ equals $\hat{\sigma}_{\mathrm{I}}$, contact detachment is triggered even at the external edge of the contact, where the pressure peak has reached its maximum value (curve C in Fig. \ref{FPmode2mode1}c).
Once again, complete detachment occurs when $\hat{\sigma}_{\mathrm{ave}} \approx \hat{\sigma}_{\mathrm{III}}$ (point D).

\subsection{Mushroom-shaped pillar - Mode III}
The presence of the plate inhibits the occurrence of Mode I \cite{carbone2011}. Therefore, in absence of interfacial defect, contact debonding will occur according to Mode III with an abrupt detachment not preceded by any progressive reduction of the contact area. 

\begin{figure}[H]
    \centering\includegraphics[width=\textwidth]{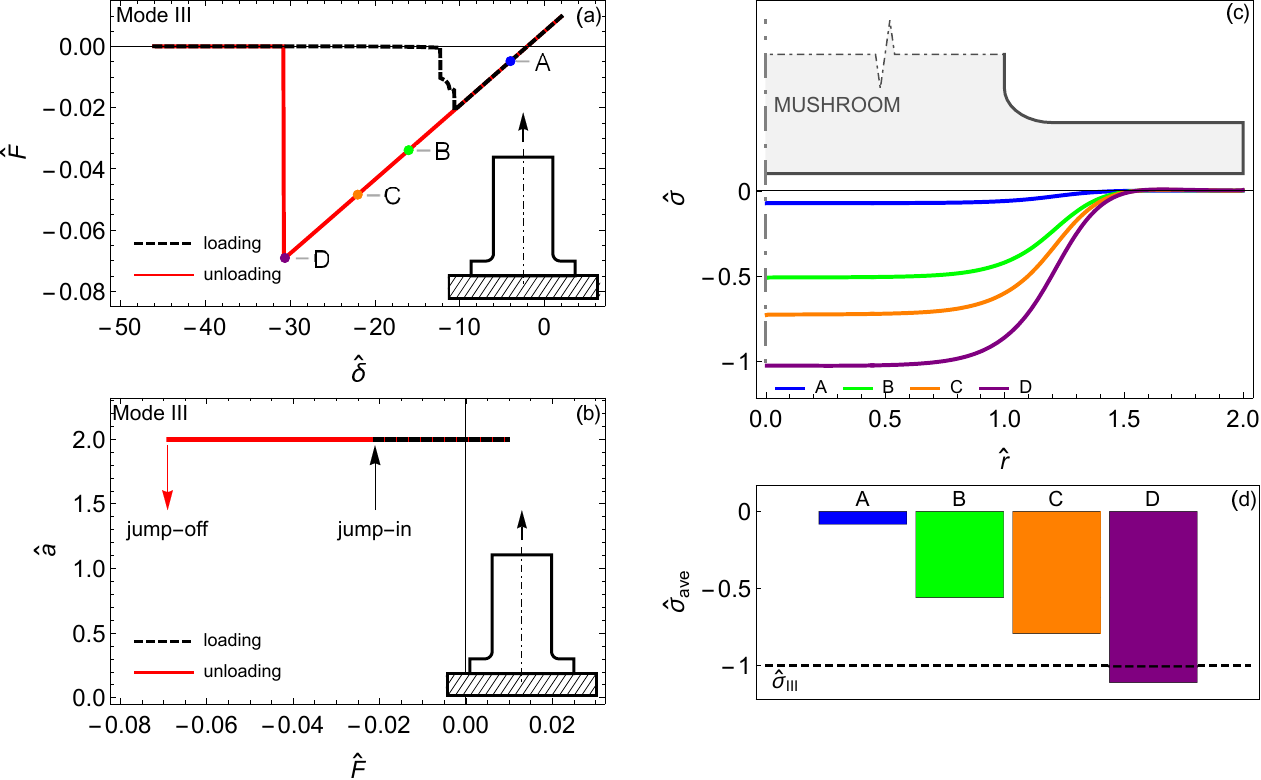}
    \caption{Mode III detachment of the mushroom-shaped pillar: a) Normalized force $\hat{F}$ plotted against the dimensionless approach $\hat{\delta}$ during the loading (black dashed line) and unloading (red solid line) phases. b) Dimensionless contact radius $\hat{a}$ shown as a function of the force $\hat{F}$. c) Interfacial pressure distribution $\hat{\sigma}$ and d) average pressure $\hat{\sigma}_\mathrm{ave}$ displayed for various values of the imposed approach $\hat{\delta}$.}
    \label{Mmode3}
\end{figure}

This scenario is confirmed by the plots of Fig. \ref{Mmode3}, with particular emphasis on Fig. \ref{Mmode3}b, where it is evident that the contact jump-off occurs without any detectable variation in the contact radius. Moreover,
Fig. \ref{Mmode3}c confirms that the pressure distribution under the pillar is almost constant, with its maximum value continuously increasing until, at the pull-off point D, the average pressure reaches the critical stress $\hat{\sigma}_{\mathrm{III}}$ (Fig. \ref{Mmode3}d).

\subsection{Mushroom-shaped Pillar - Mode II}
The presence of an inner defect ($\hat{d} = 0.446$) activates Mode II detachment being the critical stresses $\hat{\sigma}_\mathrm{II}=0.593$ and $\hat{\sigma}_\mathrm{III}=1$.

\begin{figure}[H]
    \centering\includegraphics[width=\textwidth]{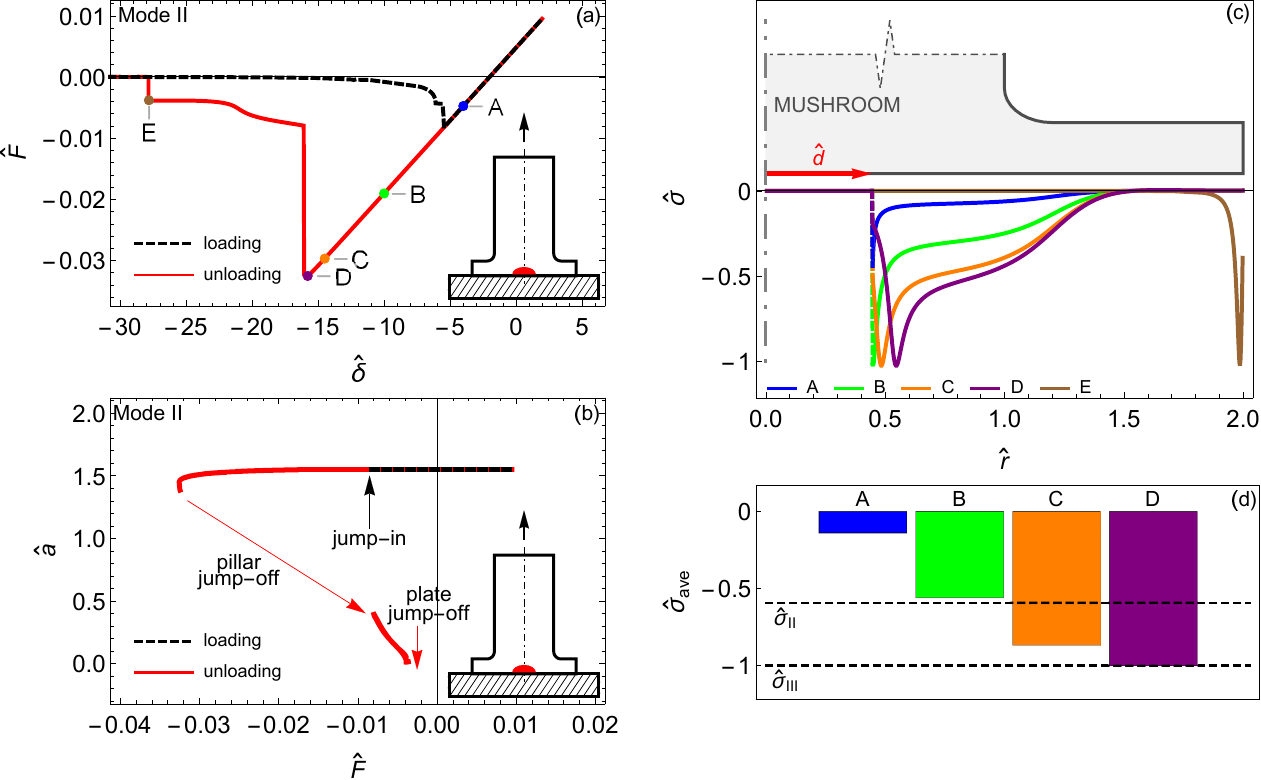}
    \caption{Mode II detachment of the mushroom-shaped pillar: a) Normalized force $\hat{F}$ plotted against the dimensionless approach $\hat{\delta}$ during the loading (black dashed line) and unloading (red solid line) phases. b) Dimensionless contact radius $\hat{a}$ shown as a function of the force $\hat{F}$. c) Interfacial pressure distribution $\hat{\sigma}$ and d) average pressure $\hat{\sigma}_\mathrm{ave}$ displayed for various values of the imposed approach $\hat{\delta}$.}
    \label{Mmode2}
\end{figure}

Figure \ref{Mmode2} distinctly illustrates two phases in the detachment process.
As the average pressure approaches the critical value $\hat{\sigma}_\mathrm{II}=0.593$ (point B), contact opening initiates at the defect edge and subsequently extends until an initial abrupt detachment takes place at pull-off point D, where $\hat{\sigma}_\mathrm{ave}=\hat{\sigma}_\mathrm{III}$.
This initial phase of debonding corresponds to the translation of the peak of pressure in Fig. \ref{FPmode1mode2}c from curve B to curve D.
However, this detachment is limited to the inner region of the pillar (as indicated by the non-vanishing force in Figs. \ref{FPmode1mode2}a and \ref{FPmode1mode2}b), and the contact remains confined under the plate. 
Advancing further into the retraction phase, the detachment continues beneath the plate until a final separation abruptly occurs at point E. This transition from pillar adhesion regime to plate adhesion regime was initially identified in Ref. \cite{afferrante2013}.

Table \ref{tabella} collects the pull-off force and hysteretic loss values for all the considered configurations. The energy loss $\hat{E}_{\mathrm{loss}}$ corresponds to the area between the loading and unloading paths in the force-penetration curve. As expected, the mushroom-shaped pillar performs better than the cylindrical one, showing a stronger adhesion capacity.

\begin{table}[h]
    \centering
    \begin{tabular}{|c|c|c|}
    \hline
    Pillar configuration & $\hat{F}_\mathrm{pull-off}$ & $\hat{E}_{\mathrm{loss}}$\\
    \hline
    cylindrical pillar (Mode I) & -0.0269 & 0.1520 \\
    \hline
    cylindrical pillar (Mode I + II) & -0.0255 & 0.1424 \\
    \hline
    cylindrical pillar (Mode II + I) & -0.0156 & 0.0576 \\
    \hline
    mushroom-shaped pillar (Mode II) & -0.0325 & 0.2731 \\
    \hline
    mushroom-shaped pillar (Mode III) & -0.0693 & 0.8881 \\
    \hline
    \end{tabular}
    \caption{Pull-off force and hysteretic loss values for the different pillar geometries and detachment modes.}
    \label{tabella}
\end{table}

\section{Conclusions}
Theoretical considerations \cite{carbone2011} enable to predict the initiation of detachment for cylindrical and mushroom-shaped pillars adhering to a rigid surface. Numerical simulations performed in the present work further contribute by characterizing the entire detachment process, spanning from the initial onset to the final contact rupture. These simulations reveal that two modes of detachment may arise during the debonding process, as a result of the redistribution of the interfacial pressures.

Independently of the detachment mode, our cohesive model suggests that the contact at a generic interfacial point is lost when the pressure at that point reaches the theoretical strength $\sigma_{\mathrm{III}}$, while complete detachment occurs when the average pressure exceeds $\sigma_{\mathrm{III}}$.

\textcolor{black}{The results show significant differences in the detachment behavior and energy dissipation between different pillar configurations and detachment modes. In this regard, we highlight that our model predict hysteretic energy loss as a result of the differences between the approach and retraction curves. Adhesion hysteresis is a phenomenon frequently observed in experiments \cite{zhang2021}, and is often overlooked in numerical and theoretical studies that typically focus solely on the detachment phase without considering the contact closure \cite{carbone2013, aksak2014}.}

\textcolor{black}{Furthermore, the presence of the plate in mushroom-shaped pillars inhibits Mode I detachment, significantly altering the adhesive performance. For example, the mushroom-shaped pillar in Mode III exhibits a hysteretic loss and a pull-off force that are approximately 484$\%$ and 158$\%$, respectively, higher than those of the cylindrical pillar in Mode I.}

\section*{Acknowledgements}
This work was  supported by the Italian Ministry of University and Research under the Programme “Department of Excellence” Legge 232/2016 (Grant No. CUP - D93C23000100001) and by the European Union - NextGenerationEU through the Italian Ministry of University and Research under the programs: PRIN2022 (Projects of Relevant National Interest) grant nr. 2022SJ8HTC - ELectroactive gripper For mIcro-object maNipulation (ELFIN); PRIN2022 PNRR (Projects of Relevant National Interest) grant nr. P2022MAZHX - TRibological modellIng for sustainaBle design Of induStrial friCtiOnal inteRfacEs (TRIBOSCORE).


\begin{thebibliography}{99}                                                                                               %


\bibitem{brodoceanu2016}
D.~Brodoceanu, C.~Bauer, E.~Kroner, E.~Arzt, T.~Kraus, Hierarchical bioinspired
  adhesive surfaces—a review, Bioinspiration \& biomimetics 11~(5) (2016)
  051001.

\bibitem{favi2014}
P.~M. Favi, S.~Yi, S.~C. Lenaghan, L.~Xia, M.~Zhang, Inspiration from the
  natural world: from bio-adhesives to bio-inspired adhesives, Journal of
  Adhesion Science and Technology 28~(3-4) (2014) 290--319.

\bibitem{purtov2015}
J.~Purtov, M.~Frensemeier, E.~Kroner, Switchable adhesion in vacuum using
  bio-inspired dry adhesives, ACS applied materials \& interfaces 7~(43) (2015)
  24127--24135.

\bibitem{autumn2002}
K.~Autumn, A.~M. Peattie, Mechanisms of adhesion in geckos, Integrative and
  comparative biology 42~(6) (2002) 1081--1090.

\bibitem{gorb2007}
S.~Gorb, M.~Varenberg, A.~Peressadko, J.~Tuma, Biomimetic mushroom-shaped
  fibrillar adhesive microstructure, Journal of The Royal Society Interface
  4~(13) (2007) 271--275.

\bibitem{autumn2002B}
K.~Autumn, M.~Sitti, Y.~A. Liang, A.~M. Peattie, W.~R. Hansen, S.~Sponberg,
  T.~W. Kenny, R.~Fearing, J.~N. Israelachvili, R.~J. Full, Evidence for van
  der waals adhesion in gecko setae, Proceedings of the National Academy of
  Sciences 99~(19) (2002) 12252--12256.

\bibitem{hosoda2021}
N.~Hosoda, M.~Nakamoto, T.~Suga, S.~N. Gorb, Evidence for intermolecular forces
  involved in ladybird beetle tarsal setae adhesion, Scientific Reports 11~(1)
  (2021) 7729.

\bibitem{zhang2015}
E.~Zhang, Y.~Wang, T.~Lv, L.~Li, Z.~Cheng, Y.~Liu, Bio-inspired design of
  hierarchical pdms microstructures with tunable adhesive superhydrophobicity,
  Nanoscale 7~(14) (2015) 6151--6158.

\bibitem{lee2018}
S.~H. Lee, S.~W. Kim, B.~S. Kang, P.-S. Chang, M.~K. Kwak, Scalable and
  continuous fabrication of bio-inspired dry adhesives with a thermosetting
  polymer, Soft Matter 14~(14) (2018) 2586--2593.

\bibitem{hensel2018}
R.~Hensel, K.~Moh, E.~Arzt, Engineering micropatterned dry adhesives: from
  contact theory to handling applications, Advanced Functional Materials
  28~(28) (2018) 1800865.

\bibitem{bullock2011}
J.~M. Bullock, W.~Federle, Beetle adhesive hairs differ in stiffness and
  stickiness: in vivo adhesion measurements on individual setae,
  Naturwissenschaften 98 (2011) 381--387.

\bibitem{kim2020}
Y.~Kim, C.~Yang, Y.~Kim, G.~X. Gu, S.~Ryu, Designing an adhesive pillar shape
  with deep learning-based optimization, ACS applied materials \& interfaces
  12~(21) (2020) 24458--24465.

\bibitem{carbone2013}
G.~Carbone, E.~Pierro, A review of adhesion mechanisms of mushroom-shaped
  microstructured adhesives, Meccanica 48 (2013) 1819--1833.

\bibitem{aksak2014}
B.~Aksak, K.~Sahin, M.~Sitti, The optimal shape of elastomer mushroom-like
  fibers for high and robust adhesion, Beilstein journal of nanotechnology
  5~(1) (2014) 630--638.

\bibitem{afferrante2013}
L.~Afferrante, G.~Carbone, The mechanisms of detachment of mushroom-s haped
  micro-p illars: From defect propagation to membrane peeling, Macromolecular
  Reaction Engineering 7~(11) (2013) 609--615.

\bibitem{micciche2014}
M.~Miccich{\'e}, E.~Arzt, E.~Kroner, Single macroscopic pillars as model system
  for bioinspired adhesives: influence of tip dimension, aspect ratio, and tilt
  angle, ACS applied materials \& interfaces 6~(10) (2014) 7076--7083.

\bibitem{paretkar2013}
D.~R. Paretkar, M.~D. Bartlett, R.~McMeeking, A.~J. Crosby, E.~Arzt, Buckling
  of an adhesive polymeric micropillar, The Journal of Adhesion 89~(2) (2013)
  140--158.

\bibitem{greiner2007}
C.~Greiner, A.~Del~Campo, E.~Arzt, Adhesion of bioinspired micropatterned
  surfaces: effects of pillar radius, aspect ratio, and preload, Langmuir
  23~(7) (2007) 3495--3502.

\bibitem{wang2014}
Y.~Wang, H.~Hu, J.~Shao, Y.~Ding, Fabrication of well-defined mushroom-shaped
  structures for biomimetic dry adhesive by conventional photolithography and
  molding, ACS applied materials \& interfaces 6~(4) (2014) 2213--2218.

\bibitem{fischer2017}
S.~C. Fischer, E.~Arzt, R.~Hensel, Composite pillars with a tunable interface
  for adhesion to rough substrates, ACS applied materials \& interfaces 9~(1)
  (2017) 1036--1044.

\bibitem{kasem2013}
H.~Kasem, M.~Varenberg, Effect of counterface roughness on adhesion of
  mushroom-shaped microstructure, Journal of The Royal Society Interface
  10~(87) (2013) 20130620.

\bibitem{afferrante2023A}
L.~Afferrante, G.~Violano, D.~Dini, How does roughness kill adhesion?, Journal
  of the Mechanics and Physics of Solids (2023) 105465.

\bibitem{afferrante2023B}
L.~Afferrante, G.~Violano, G.~Carbone, Exploring the dynamics of viscoelastic
  adhesion in rough line contacts, Scientific Reports 13~(1) (2023) 15060.

\bibitem{carbone2011}
G.~Carbone, E.~Pierro, S.~N. Gorb, Origin of the superior adhesive performance
  of mushroom-shaped microstructured surfaces, Soft Matter 7~(12) (2011)
  5545--5552.

\bibitem{zhang2021}
X.~Zhang, Y.~Wang, R.~Hensel, E.~Arzt, A design strategy for mushroom-shaped
  microfibrils with optimized dry adhesion: experiments and finite element
  analyses, Journal of Applied Mechanics 88~(3) (2021) 031015.

\bibitem{CarboneSMALL}
E.~P. Giuseppe~Carbone, Sticky bio-inspired micropillars: Finding the best
  shape, SMALL 8~(9) (2012) 1449--1454.

\bibitem{rey2017}
V.~Rey, G.~Anciaux, J.-F. Molinari, Normal adhesive contact on rough surfaces:
  efficient algorithm for fft-based bem resolution, Computational Mechanics 60
  (2017) 69--81.

\bibitem{salehani2019}
M.~K. Salehani, N.~Irani, L.~Nicola, Modeling adhesive contacts under
  mixed-mode loading, Journal of the Mechanics and Physics of Solids 130 (2019)
  320--329.

\bibitem{popov2021}
V.~L. Popov, Q.~Li, I.~A. Lyashenko, R.~Pohrt, Adhesion and friction in hard
  and soft contacts: Theory and experiment, Friction 9 (2021) 1688--1706.

\bibitem{sanner2022}
A.~Sanner, L.~Pastewka, Crack-front model for adhesion of soft elastic spheres
  with chemical heterogeneity, Journal of the Mechanics and Physics of Solids
  160 (2022) 104781.

\bibitem{afferrante2022}
L.~Afferrante, G.~Violano, On the effective surface energy in viscoelastic
  hertzian contacts, Journal of the Mechanics and Physics of Solids 158 (2022)
  104669.

\bibitem{violano2022A}
G.~Violano, L.~Afferrante, Size effects in adhesive contacts of viscoelastic
  media, European Journal of Mechanics-A/Solids 96 (2022) 104665.

\bibitem{carbone2012}
G.~Carbone, E.~Pierro, Effect of interfacial air entrapment on the adhesion of
  bio-inspired mushroom-shaped micro-pillars, Soft Matter 8~(30) (2012)
  7904--7908.

\bibitem{kossa2023}
A.~Kossa, R.~Hensel, R.~M. McMeeking, Adhesion of a cylindrical punch with
  elastic properties that vary radially, Mechanics Research Communications 130
  (2023) 104123.

\bibitem{maugis2000}
D.~Maugis, Contact, adhesion and rupture of elastic solids, Vol. 130, Springer
  Science \& Business Media, 2000.

\bibitem{johnson1971}
K.~L. Johnson, K.~Kendall, a.~Roberts, Surface energy and the contact of
  elastic solids, Proceedings of the royal society of London. A. mathematical
  and physical sciences 324~(1558) (1971) 301--313.

\bibitem{violano2019}
G.~Violano, L.~Afferrante, Modeling the adhesive contact of rough soft media
  with an advanced asperity model, Tribology Letters 67 (2019) 1--7.

\bibitem{lorenz2013}
B.~Lorenz, B.~Krick, N.~Mulakaluri, M.~Smolyakova, S.~Dieluweit, W.~Sawyer,
  B.~Persson, Adhesion: role of bulk viscoelasticity and surface roughness,
  Journal of Physics: Condensed Matter 25~(22) (2013) 225004.

\bibitem{violano2021rateA}
G.~Violano, A.~Chateauminois, L.~Afferrante, Rate-dependent adhesion of
  viscoelastic contacts, part i: Contact area and contact line velocity within
  model randomly rough surfaces, Mechanics of Materials 160 (2021) 103926.

\bibitem{violano2019B}
G.~Violano, L.~Afferrante, Adhesion of compliant spheres: An experimental
  investigation, Procedia Structural Integrity 24 (2019) 251--258.

\bibitem{PR2014}
L.~Pastewka, M.~O. Robbins, Contact between rough surfaces and a criterion for
  macroscopic adhesion, Proceedings of the National Academy of Sciences 111~(9)
  (2014) 3298--3303.

\bibitem{aksak2011RATIO}
B.~Aksak, C.-Y. Hui, M.~Sitti, The effect of aspect ratio on adhesion and
  stiffness for soft elastic fibres, Journal of The Royal Society Interface
  8~(61) (2011) 1166--1175.

\bibitem{delcampo2007ALTEZZA}
A.~Del~Campo, C.~Greiner, E.~Arzt, Contact shape controls adhesion of
  bioinspired fibrillar surfaces, Langmuir 23~(20) (2007) 10235--10243.

\bibitem{muller1983}
V.~Muller, B.~Derjaguin, Y.~P. Toporov, On two methods of calculation of the
  force of sticking of an elastic sphere to a rigid plane, Colloids and
  Surfaces 7~(3) (1983) 251--259.

\bibitem{violano2021HYST}
G.~Violano, L.~Afferrante, Roughness-induced adhesive hysteresis in self-affine
  fractal surfaces, Lubricants 9~(1) (2021) 7.

\bibitem{wang2021}
A.~Wang, Y.~Zhou, M.~H. M{\"u}ser, Modeling adhesive hysteresis, Lubricants
  9~(2) (2021) 17.

\bibitem{feng2001}
J.~Q. Feng, Adhesive contact of elastically deformable spheres: a computational
  study of pull-off force and contact radius, Journal of colloid and interface
  science 238~(2) (2001) 318--323.
  
\end{thebibliography}
\end{document}